\documentclass[review,3p]{elsarticle}

\usepackage{graphicx}
\usepackage{epsfig}
\usepackage{color}
\usepackage{psfrag}
\usepackage{subfigure}
\usepackage{subfigmat}
\usepackage{subeqnarray}
\usepackage{amsmath}
\usepackage{amsfonts,stmaryrd,amssymb,theorem}
\usepackage{xcolor,colortbl}
\usepackage[mathscr]{euscript}
\usepackage[ruled]{algorithm2e}
\usepackage{algorithmic}
\usepackage{changepage}
\usepackage{multirow}
\graphicspath{{../Figures/}{./Figure/}}
\usepackage{rotating}
\usepackage{bm}
\usepackage{lineno}
\usepackage{textgreek}
\usepackage{enumitem}
\usepackage{array}
\usepackage{tikz}

\usepackage{booktabs}
\usepackage{stackengine}

\usepackage[export]{adjustbox}
\usepackage{epstopdf}
\usepackage[capitalize]{cleveref}

\journal{AIAA SciTech 2018, Kissimmee, FL}

\makeatletter
\newcommand\rlarrows{\mathop{\operator@font \rightleftarrows}\nolimits}
\makeatother
\DeclareFontFamily{OT1}{pzc}{}
\DeclareFontShape{OT1}{pzc}{m}{it}{<-> s * [1.10] pzcmi7t}{}
\DeclareMathAlphabet{\mathpzc}{OT1}{pzc}{m}{it}

 \def\0vec{{\mbox{\boldmath$0$}}}

\begin{document}

%%%%%%%%%%%%%%%%%%%%%%%%%%%%%%%%%%%%%%%%%%
\title{Numerical analysis on mixing processes for transcritical real-fluid simulations}

\author[Stanford]{Peter C. Ma\corref{cor}}
\ead{peterma@stanford.edu}
\author[Stanford]{Hao Wu}
%\ead{wuhao@stanford.edu}
\author[Stanford]{Daniel T. Banuti}
%\ead{dbanuti@stanford.edu}
\author[Stanford]{Matthias Ihme}
%\ead{mihme@stanford.edu}

\address[Stanford]{Department of Mechanical Engineering, Stanford University, Stanford, CA 94305, USA}

\cortext[cor]{Corresponding author}
%%%%%%%%%%%%%%%%%%%%%%%%%%%%%%%%%%%%%%%%%%

%%%%%%%%%%%%%%%%%%%%%%%%%%%%%%%%%%%%%%%%%%
\begin{abstract}
The accurate and robust simulation of transcritical real-fluid flows is crucial for many engineering applications. Diffused interface methods are frequently employed and several numerical schemes have been developed for simulating transcritical flows. These schemes can be categorized into two types, namely fully conservative and quasi-conservative schemes. An adaptive scheme which is a hybrid of the two is developed in this study. By considering several numerical test cases, it is shown that different schemes predict distinctly different mixing behaviors. It is shown that the mixing processes follow the isobaric-adiabatic and isobaric-isochoric mixing models for fully and quasi-conservative schemes, respectively, and the adaptive scheme yields a mixing behavior that spans both models. The distinct mixing behaviors are a consequence of numerical diffusion instead of physical diffusion and can be attributed to insufficient numerical spatial resolution. This work provides a better understanding on the interpretation of numerical simulation results and the mixing models that are commonly used to study transcritical flows. 
\end{abstract}
%%%%%%%%%%%%%%%%%%%%%%%%%%%%%%%%%%%%%%%%%%

\maketitle

%=====================================================================
\section{Introduction}
%=====================================================================
% transcritical injection important, phase separation may occur but we don't know for sure
Transcritical injection is used widely in diesel engines, gas turbines, and rocket engines~\cite{mayer2000injection, oschwald2006injection, chehroudi2012recent, Oefelein2012}. In accordance with pure fluid behavior, it has been assumed traditionally that fluids at supercritical pressures do not break up into droplets upon injection, but disintegrate through a turbulent mixing process~\cite{oschwald2006injection, Newman1971, Mayer1996, Candel2006}. However, from a thermodynamic point of view, the mixture critical pressure may significantly exceed the critical pressures of the pure components~\cite{Elliott2012}. Then, at a chamber pressure that is nominally supercritical with respect to the pure propellants, the local mixture may experience a subcritical pressure, allowing for phase separation~\cite{poling2001properties, Habiballah2006, dahms2015non}. This has been observed experimentally for different fluid mixtures~\cite{Mayer1998, Roy2013, Manin2014}. However, determining whether or not phase separation may occur under typical transcritical conditions, and the importance of interfaces and surface tension forces under these high pressures, remain open questions~\cite{yang2000modeling, bellan2000supercritical, Dahms2013a, banuti2017seven}.

% different numerical schemes have been developed
To study transcritical flows numerically, diffused interface methods have been widely used and different numerical schemes have been adopted~\cite{schmitt2010large, ruiz2012unsteady, terashima2012approach, matheis2016multi, hickey2013large, kawai2015robust, lacaze2017robust, pantano2017oscillation, yang2017comparison, unnikrishnan2017subgrid}. The surface tension force is typically not considered in these solvers. Traditionally, fully conservative (FC) schemes have been used for transcritical flows. However, several groups reported numerical difficulties or even failures with FC schemes in conjunction with a real-fluid state equation, due to the occurrence of spurious pressure oscillations~\cite{schmitt2010large, matheis2016multi, hickey2013large, lacaze2017robust}. This has motivated the development of quasi-conservative (QC) schemes for transcritical flows. Schmitt~et~al.~\cite{schmitt2010large, ruiz2012unsteady} added a correction term in the energy equation by connecting the artificial dissipation terms in the mass, momentum, and energy conservation equations and setting the pressure differential to zero. Since the correction term is not in flux form, the scheme is not strictly energy-conservative. Terashima, Kawai and coworkers~\cite{terashima2012approach, kawai2015robust} solved a transport equation for pressure instead of the total energy equation in their finite difference solver so that the pressure equilibrium across contact interfaces is maintained. Pantano~et~al.~\cite{pantano2017oscillation} formulated a numerical scheme for transcritical contact and shock problems, which introduces an additional non-conservative transport equation for maintaining the mechanical equilibrium of pressure. Ma~et~al.~\cite{MaJCP2017, ma2014supercritical} extended a double-flux model to the transcritical regime, and later applied the model for transcritical combustion~\cite{ma2017framework}.

% different mixing behaviors
Discrepancies in the mixing behaviors have been reported between FC and QC schemes in the literature. The solver AVBP with a QC scheme and the FC solver RAPTOR were both used to simulate a benchmark test case of a LOX/GH2 mixing layer~\cite{ruiz2015numerical}. Despite similar behaviors of axial and transverse velocity profiles and species mass fractions, significant differences in both the mean and root-mean-square (rms) values of the temperature field were observed. The origin of these differences were not investigated. Large-eddy simulations (LES) of the ECN Spray A case~\cite{pickett2011engine} were performed by Matheis~and~Hickel~\cite{matheis2016multi}, and the temperature field was found to be significantly higher when using a QC scheme, which was attributed to the energy conservation error. Lacaze~et~al.~\cite{lacaze2017robust} pointed out that the adiabatic mixing profile can be obtained only by using a FC scheme. To fully understand the impact of numerical methods, a systematic analysis is required in which the distinct mixing behaviors between the two types of numerical schemes are investigated.

% adiabatic mixing used as a golden standard
The adiabatic mixing process has been considered in several studies to describe transcritical flows. For numerical simulations, Lacaze~et~al.~\cite{lacaze2015analysis} compared their simulation results with the adiabatic mixing and attribute the slight deviations of the solutions from adiabatic mixing to the transport anomalies. For theoretical analysis, Dahms~and~Oefelein~\cite{dahms2015non,Dahms2013a} approached the problem using linear gradient theory in addition to the adiabatic mixing assumption. Qiu~and~Reitz~\cite{Qiu2015} focused on the thermodynamic stability of the local mixture, accounting for a subsequent reduction in temperature when a phase separation occurs. Experimentally, the ECN workshop~\cite{pickett2011engine} has been using adiabatic mixing to transform the mixture fraction measurements into the temperature field for the validation of numerical simulations.

% methods not assuming adiabatic mixing
However, there are a number of studies in which diffusion of heat and mass are treated individually, without imposing additional assumptions. The thermodynamic behavior of isolated transcritical droplets in gaseous environments has been studied computationally~\cite{Delplanque1993, Yang1994, Haldenwang1996, Harstad1998, Sirignano1999}, typically relying on a transition from a sharp to a diffused interface when the mixture critical temperature is locally exceeded. Analyzing instead counterflow diffusion flames, the local thermodynamic state in hydrogen-oxygen combustion was evaluated~\cite{Banuti2016sub, Lacaze2012} using classical mixing rules~\cite{poling2001properties}, showing that reactive cases are more susceptible to phase separation than inert mixing, as the product water significantly reduces the local mixture critical pressure. The $Y$-$T$ mixing lines, shown by Harstad~and~Bellan~\cite{Harstad1998}, deviate qualitatively from the adiabatic mixing lines, which stands in contrast to the simultaneous change in both mole fraction and temperature observed during adiabatic mixing. Harstad~and~Bellan~\cite{Harstad1999} attribute this to an effective Lewis number in the case of transcritical evaporation, while adiabatic mixing corresponds to a unity Lewis number assumption. Banuti~et~al.~\cite{BanutiCnF2016} pointed out that break-up of the liquid oxygen jet in LOX/GH2 injection similarly occurs under essentially pure fluid conditions, in a process in which heat is transferred while diffusion is suppressed.

% objectives and outline
The objective of this study is to investigate the representation of mixing processes under transcritical conditions from different numerical schemes through numerical analysis and simulations. Governing equations, thermodynamic relations and numerical schemes used in this study are presented in \cref{sec:math}. Test cases under typical transcritical injection conditions are conducted in \cref{sec:results}, and the mixing behaviors are investigated by comparing the solutions among different methods. Two mixing types are identified. The paper finishes with conclusions in \cref{sec:conclusions}.

%=====================================================================
\section{Mathematical Formulations}\label{sec:math}
%=====================================================================

%=====================================================================
\subsection{Governing equations}\label{sec:governing}
%=====================================================================
The governing equations for the diffused interface method considered in this study are the conservation of mass, momentum, total energy, and species, which take the following form
\begin{subequations}
    \label{eqn:governingEqn}
    \begin{align}
        \frac{\partial \rho}{\partial t} +\nabla \cdot (\rho \boldsymbol{u}) &= 0\,,\\
        \frac{\partial (\rho \boldsymbol{u})}{\partial t} + \nabla \cdot (\rho \boldsymbol{u} \boldsymbol{u} + p \boldsymbol{I}) &= \nabla \cdot \boldsymbol{\tau} \,,\\
        \frac{\partial  (\rho E)}{\partial t} + \nabla \cdot [\boldsymbol{u}(\rho E + p)] &= \nabla \cdot (\boldsymbol{\tau} \cdot \boldsymbol{u}) -\nabla \cdot \boldsymbol{q} \,,\\
        \frac{\partial (\rho Y_k)}{\partial t} + \nabla \cdot (\rho \boldsymbol{u} Y_k) &= \nabla \cdot (\rho D_k \nabla Y_k)  \;\; \text{for}\;\; k = 1\,, \dots, N_S -1 \,,
    \end{align}
\end{subequations}
where $\rho$ is the density, $\boldsymbol{u}$ is the velocity vector, $p$ is the pressure, $E$ is the specific total energy, $Y_k$ is the mass fraction of species $k$, $D_k$ is the diffusion coefficient for species $k$, and $N_S$ is the number of species. The viscous stress tensor and heat flux are written as
\begin{subequations}
    \begin{align}
        &\boldsymbol{\tau} =  \mu \left[ \nabla \boldsymbol{u} + (\nabla \boldsymbol{u})^T \right] -\frac{2}{3}\mu (\nabla \cdot \boldsymbol{u}) \boldsymbol{I} \;,\\
        &\boldsymbol{q} = - \lambda \nabla T - \rho \sum_{k = 1}^{N_S} {h}_k D_k \nabla Y_k\;,
    \end{align}
\end{subequations}
where $T$ is the temperature, $\mu$ is the dynamic viscosity, $\lambda$ is the thermal conductivity, and ${h}_k$ is the partial enthalpy of species $k$. The total energy is related to the internal energy and the kinetic energy
\begin{equation}
    \rho E = \rho e + \frac{1}{2}\rho \boldsymbol{u} \cdot \boldsymbol{u}\,.
\end{equation}
The system is closed with an equation of state (EoS), which is here written in pressure-explicit form as
\begin{equation}
    \label{eqn:eos}
    p = f(\rho, e, \boldsymbol{Y})\,,
\end{equation}
where $\boldsymbol{Y} = [Y_1, \dots, Y_k]^T$ represents the species vector.

%=====================================================================
\subsection{Thermodynamic relations}\label{sec:eos}
%=====================================================================

%-----------------------------------------------------------------
\begin{table}[!b!]
    \centering
    \begin{tabular}{c c c c c c c}
        \toprule 
        Species & $W$ [kg/kmol] & $T_c$ [K] & $p_c$ [MPa] & $\rho_c$ [kg/m$^3$] & $Z_c$ & $\omega$\\
        \midrule
        H$_2$ & 2.02 & 33.15 & 1.30 & 31.26 & 0.303 & -0.219\\
%        N$_2$ & 28.0 & 126.2 & 3.40 & 313.3 & 0.289 & 0.0372\\
        O$_2$ & 32.0 & 154.6 & 5.04 & 436.1 & 0.287 & 0.0222\\
%        {\it n}-C$_{12}$H$_{26}$  & 170.3 & 658.1 & 1.82 & 226.5 & 0.249 & 0.574\\
%        FK-5-1-12  & 316.1 & 441.8 & 1.86 & 639.1 & 0.251 & 0.471\\
        \bottomrule
    \end{tabular}
    \caption{Thermodynamic properties (molecular weight, critical properties for temperature, pressure, density, and compressibility, and acentric factor) for different species considered in this study.\label{tab:criticalProperty}}
\end{table}
%-----------------------------------------------------------------

% PR-EoS
The analysis and results in this study are not limited to a single EoS, and are general to real-fluid transcritical flows. For computational efficiency and prevailing usage, the Peng-Robinson (PR) cubic EoS~\cite{poling2001properties, peng1976new} is used in this study, which can be written as
\begin{equation}
    \label{eqn:preos}
    p = \frac{R T}{v - b} - \frac{a}{v^2+2bv-b^2}\,,
\end{equation}
where $R$ is the gas constant, $v = 1/\rho$ is the specific volume, and the parameters $a$ and $b$ are dependent on temperature and composition to account for effects of intermolecular forces. For mixtures, the mixing rules due to Harstad~et~al.~\cite{harstad1997efficient} are used and procedures for evaluating thermodynamic quantities using PR-EoS can be found in our previous works~\cite{MaJCP2017}. Vapor liquid equilibrium (VLE) calculations are also conducted using PR-EoS and details can be found in Elliot~and~Lira~\cite{Elliott2012} and our previous studies~\cite{raju2017widom}. The thermodynamic properties of different species considered in this study are listed in \cref{tab:criticalProperty}.

%=====================================================================
\subsection{Numerical methods}\label{sec:numerics}
%=====================================================================
The unstructured solver, CharLES$^x$, is employed in this study~\cite{khalighi2011unstructured,wu2017mvp}. 
A finite volume approach is utilized for the discretization of the system of equations, \cref{eqn:governingEqn}
\begin{equation}
    \label{eqn:fv}
    \frac{\partial \boldsymbol{U}}{\partial t} V_{cv} + \sum_f (F^e_f - F^v_f) A_f = 0\,,
\end{equation}
where 
\begin{equation}
    \boldsymbol{U} = \left[\rho, (\rho \boldsymbol{u})^T, \rho E, (\rho \boldsymbol{Y})^T \right]^T
\end{equation}
is the vector of conserved variables, $F^e$ is the face-normal Euler flux vector, $F^v$ is the face-normal viscous flux vector which corresponds to the right-hand side (RHS) of \cref{eqn:governingEqn}, $V_{cv}$ is the volume of the control volume, and $A_f$ is the face area.

A Strang-splitting scheme~\cite{strang1968construction} is applied in this study to separate the convection and diffusion operators. The convective flux is discretized using a sensor-based hybrid scheme in which a high-order, non-dissipative scheme is combined with a low-order, dissipative scheme to minimize the numerical dissipation~\cite{MaJCP2017, khalighi2011unstructured}. A central scheme which is fourth-order on uniform meshes is used along with a second-order ENO scheme for the hybrid scheme. A density sensor~\cite{hickey2013large, lv2017underresolved} is adopted in this study. Due to the large density gradients under transcritical conditions, an entropy-stable flux correction technique~\cite{MaJCP2017} is used to ensure the physical realizability of the numerical solutions and to dampen non-linear instabilities in the numerical schemes. A strong stability preserving 3rd-order Runge-Kutta (SSP-RK3) scheme \cite{gottlieb2001strong} is used for time integration.

%=====================================================================
\subsection{Fully conservative (FC) and quasi-conservative (QC) schemes}\label{sec:FCQC}
%=====================================================================
Three different schemes are considered in this study for transcritical flow simulations, which are the FC scheme, the QC scheme, and a newly developed adaptive scheme that belongs to the QC schemes. The three schemes differ from each other for the treatment of the convection operators in \cref{eqn:governingEqn}, which will be discussed in the following.

\subsubsection{Fully conservative scheme}
For the FC scheme, the Euler flux of faces of cell $i$ is evaluated as
\begin{equation}
    F^e = \hat{F} (\boldsymbol{U}_j)\;\; \text{for} \;\; j \in \boldsymbol{S}_i\,,
\end{equation}
where $\hat{F}$ is the numerical flux, and $\boldsymbol{S}_i$ is the spatial stencil of cell $i$. The evaluation of numerical flux $\hat{F}$ involves spatial reconstruction and flux calculations, which are not the focus of this study. The conservative variables are updated using \cref{eqn:fv}, after which the primitive variables are calculated using the updated conservative variables. Thermodynamic variables are updated using the specified EoS, and specifically the pressure is updated by \cref{eqn:eos}, which is generally an iterative process for real-fluid EoSs.

The Euler flux evaluated for the two neighboring cells of the face is exactly the same in the FC scheme, which ensures strict conservation properties of the flow variables. However, due to the strong non-linearity inherent in the real-fluid EoS, spurious pressure oscillations will be generated when a FC scheme is used~\cite{terashima2012approach, MaJCP2017, lacaze2017robust}, which motivated the development of QC schemes.

\subsubsection{Quasi-conservative scheme}
In this study, the double-flux model~\cite{MaJCP2017, ma2017numerical} is selected as a representative QC scheme. The relation between the pressure and the internal energy is frozen in both space and time, which converts the local system to an equivalently calorically prefect gas system. This treatment removes not only the spurious pressure oscillations but also the oscillations in other physical quantities induced by the pressure oscillation. Only an outline of this method is presented here and the details are developed in \cite{MaJCP2017}. In this model, an effective specific heat ratio or adiabatic exponent, and an effective reference energy value are used to relate the pressure and the internal energy, which are defined as
\begin{subequations}
    \label{eqn:gammaS}
    \begin{align}
        &\gamma^* = \frac{\rho c^2}{p}\,,\\
        &e_0^* = e - \frac{pv}{\gamma^* - 1}\,,
    \end{align}
\end{subequations}
where $c$ is the speed of sound. The Euler flux for cell $i$ is evaluated as
\begin{equation}
    F^e = \hat{F} (\boldsymbol{U}^*_j)\;\; \text{for} \;\; j \in \boldsymbol{S}_i\,,
\end{equation}
where
\begin{equation}
     \boldsymbol{U}^* = \left[\rho, (\rho \boldsymbol{u})^T, \rho E^*, (\rho \boldsymbol{Y})^T \right]^T
\end{equation}
with
\begin{equation}
    (\rho E)^*_j = \frac{p_j}{\gamma^*_i - 1} + \rho_j e_{0,i}^* + \frac{1}{2} \rho_j \boldsymbol{u}_j \cdot \boldsymbol{u}_j\,.
\end{equation}
After the conservative variables are updated using \cref{eqn:fv}, the pressure is calculated as
\begin{equation}
    p_i = (\gamma^*_i - 1) \left[(\rho E)_i - \rho e_{0,i}^* - \frac{1}{2} \rho_i \boldsymbol{u}_i \cdot \boldsymbol{u}_i \right]\,,
\end{equation}
with other thermodynamic quantities updated using the pressure. Finally, after all the RK sub-steps, to ensure the thermodynamic consistency, the internal energy is updated from the primitive variables using \cref{eqn:eos}.

Since different frozen values of $\gamma^*$ and $e_0^*$ are used for each cell, the two energy fluxes at a face are no longer the same, yielding a QC scheme. The conservation error in total energy was shown to converge to zero with increasing resolution~\cite{MaJCP2017}. Note that the double-flux method has similar performance and conservation behavior as the pressure evolution method, which is another commonly used QC scheme for transcritical flows.

\subsubsection{Adaptive quasi-conservative scheme}
An adaptive double-flux model is also implemented and tested in this study. In this model, the double-flux method is only applied in the region where real-fluid effects are present, and for the ideal-gas part of the flow, the solver will switch back to the FC scheme. The implementation is straightforward thanks to the resemblance of the governing equations in the double-flux model to those in the FC scheme. A sensor which is based on the adiabatic exponent, $\gamma^*$, is used to identify the region where the double-flux model is needed, expressed as
\begin{equation}
    |\gamma^* - \gamma^*_\text{nb}| > \epsilon_{\gamma^*}\,,
    \label{eqn:ddf_sensor}
\end{equation}
where ``nb" refers to the neighbor cells, and $\epsilon_{\gamma^*}$ is a threshold value. This is due to the fact that the adiabatic exponent has a sharp gradients across the pseudo-boiling region where pressure oscillations are mostly severe~\cite{MaJCP2017}. The scheme with adaptive double-flux model results in an adaptive hybrid scheme between the FC and QC schemes, and in general this scheme belongs to the QC schemes.

Note that the numerics along with other thermo-transport models are kept identical for the FC, QC and adaptive schemes considered in this study. The only exceptions are the convective flux evaluation and the way how pressure is updated, which facilitates a fair comparison among different types of schemes.

%=====================================================================
\section{Results and Discussion}\label{sec:results}
%=====================================================================
Test cases that are relevant for rocket applications under transcritcial conditions are examined in this section using the numerical methods introduced in the previous sections. Simulation results obtained from these schemes will be used for the analysis of different mixing behaviors.

\subsection{LOX/GH2 mixing layer\label{sec:H2O2}}
%=========================================================================
% case description
The configuration considered here is relevant for rocket engines, where a two-dimensional mixing layer of liquid oxygen (LOX) and gaseous hydrogen (GH2) is simulated. This case was proposed by Ruiz~et~al.~\cite{ruiz2015numerical} as a benchmark case to test numerical solvers for high-Reynolds number turbulent flows with large density ratios. The LOX stream is injected at a temperature of 100~K, and GH2 is injected at a temperature of 150~K. The pressure is set to 10~MPa. The GH2 and LOX jets have velocities of 125~m/s, and 30~m/s, respectively. For more details about the operating conditions of the benchmark case, the reader is referred to Ruiz~et~al.~\cite{ruiz2015numerical}.

%-----------------------------------------------------------------
\begin{figure}[!tb!]
    \centering
    \def\stackalignment{l}
    % density
    \subfigure{
        \topinset
        {\includegraphics[height=0.04\columnwidth,clip=]{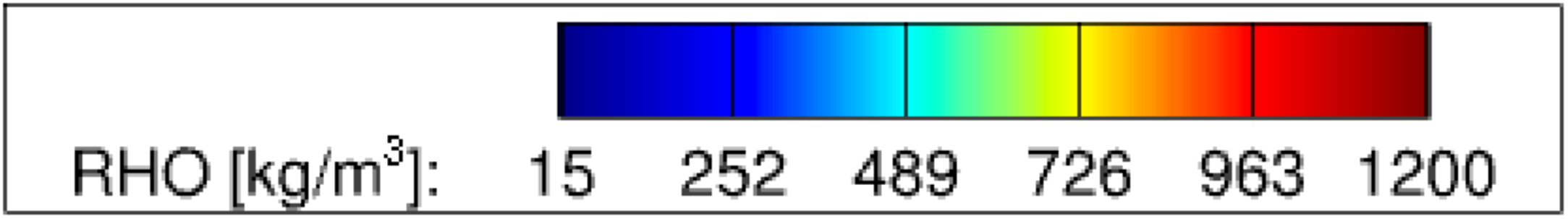}}
        {\includegraphics[width=0.475\columnwidth,trim={1cm 6cm 8.5cm 4cm},clip=]{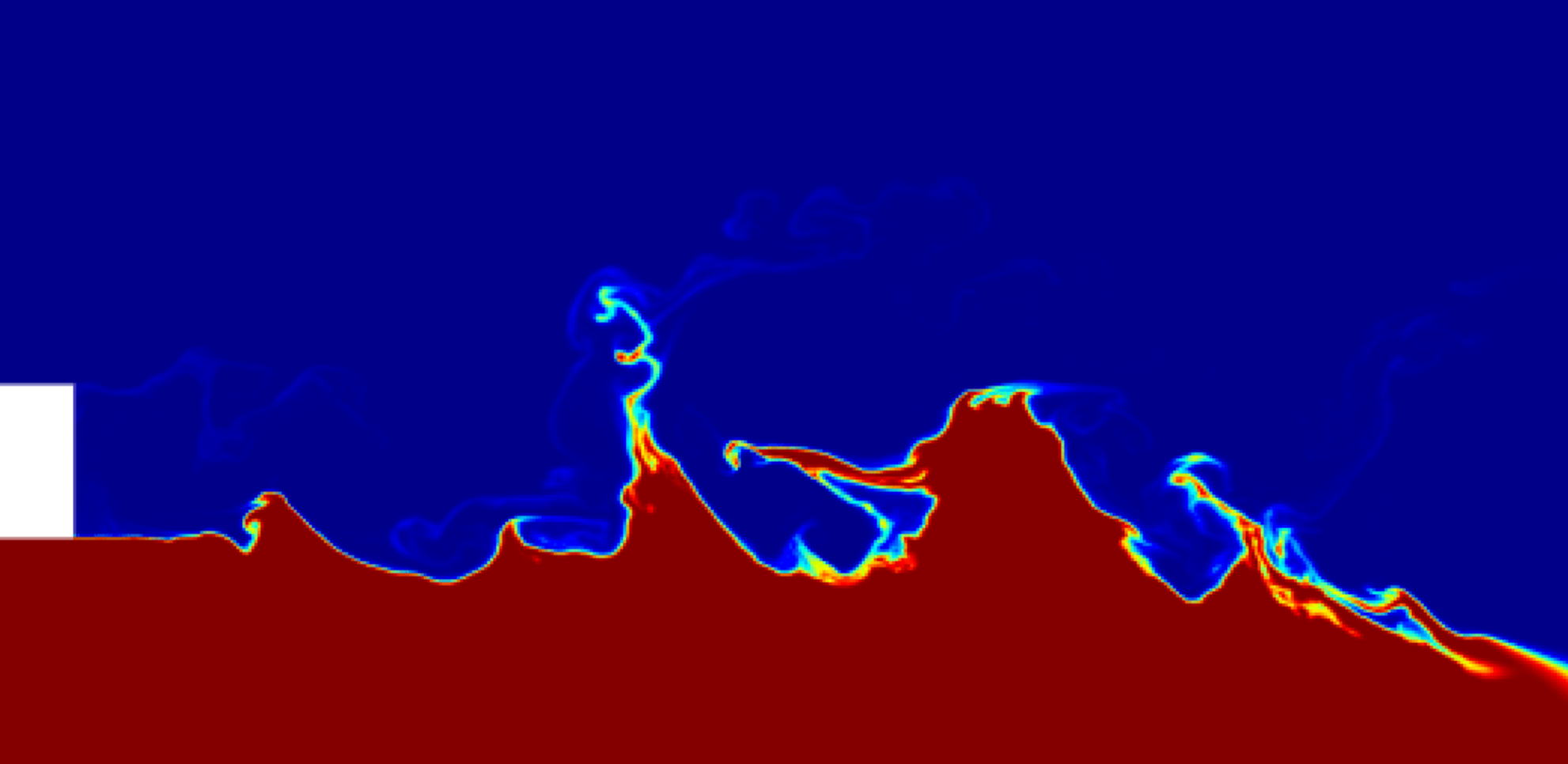}}
        {1.5pt}{1.5pt}
        \label{fig:subfig1} 
    }
    \subfigure{
        \topinset
        {\includegraphics[height=0.04\columnwidth,clip=]{rho_legend.pdf}}
        {\includegraphics[width=0.475\columnwidth,trim={1cm 6cm 8.5cm 4cm},clip=]{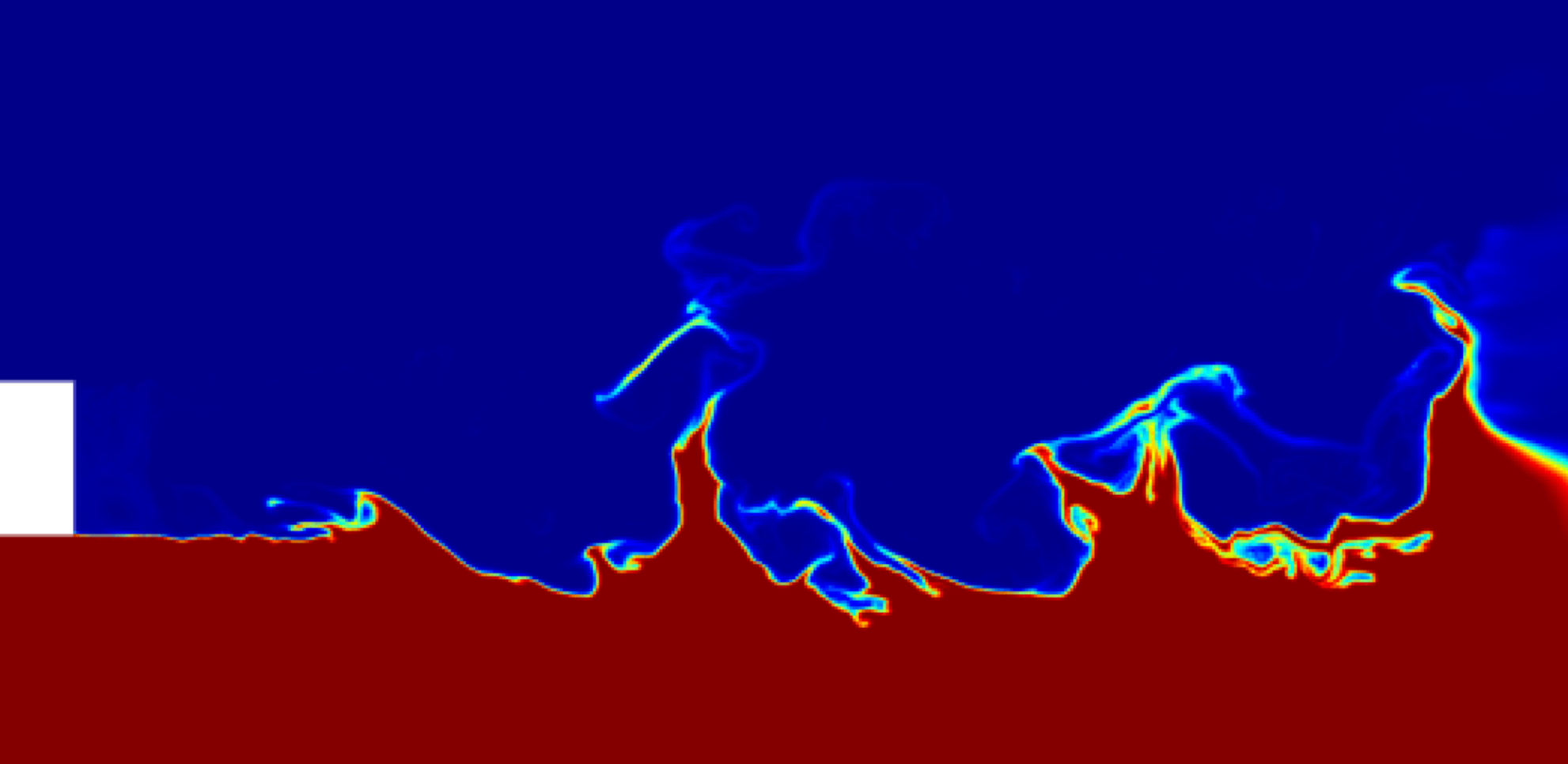}}
        {1.5pt}{1.5pt}
        \label{fig:subfig1} 
        \label{fig:subfig1} 
    }\\[-0.1em]
    % pressure
    \subfigure{
        \topinset
        {\includegraphics[height=0.04\columnwidth,clip=]{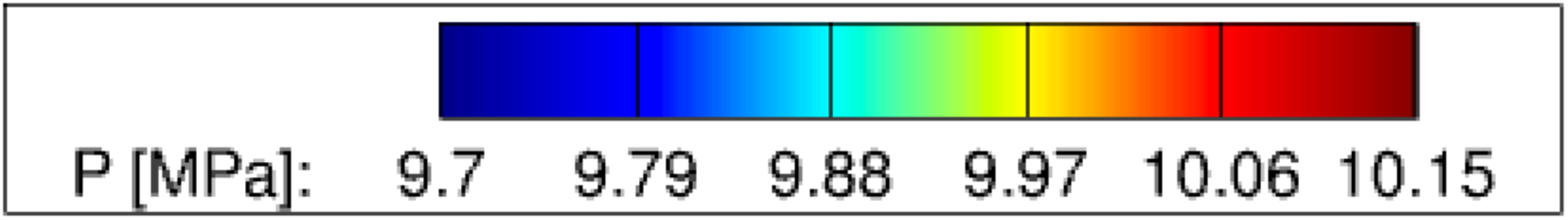}}
        {\includegraphics[width=0.475\columnwidth,trim={1cm 6cm 8.5cm 4cm},clip=]{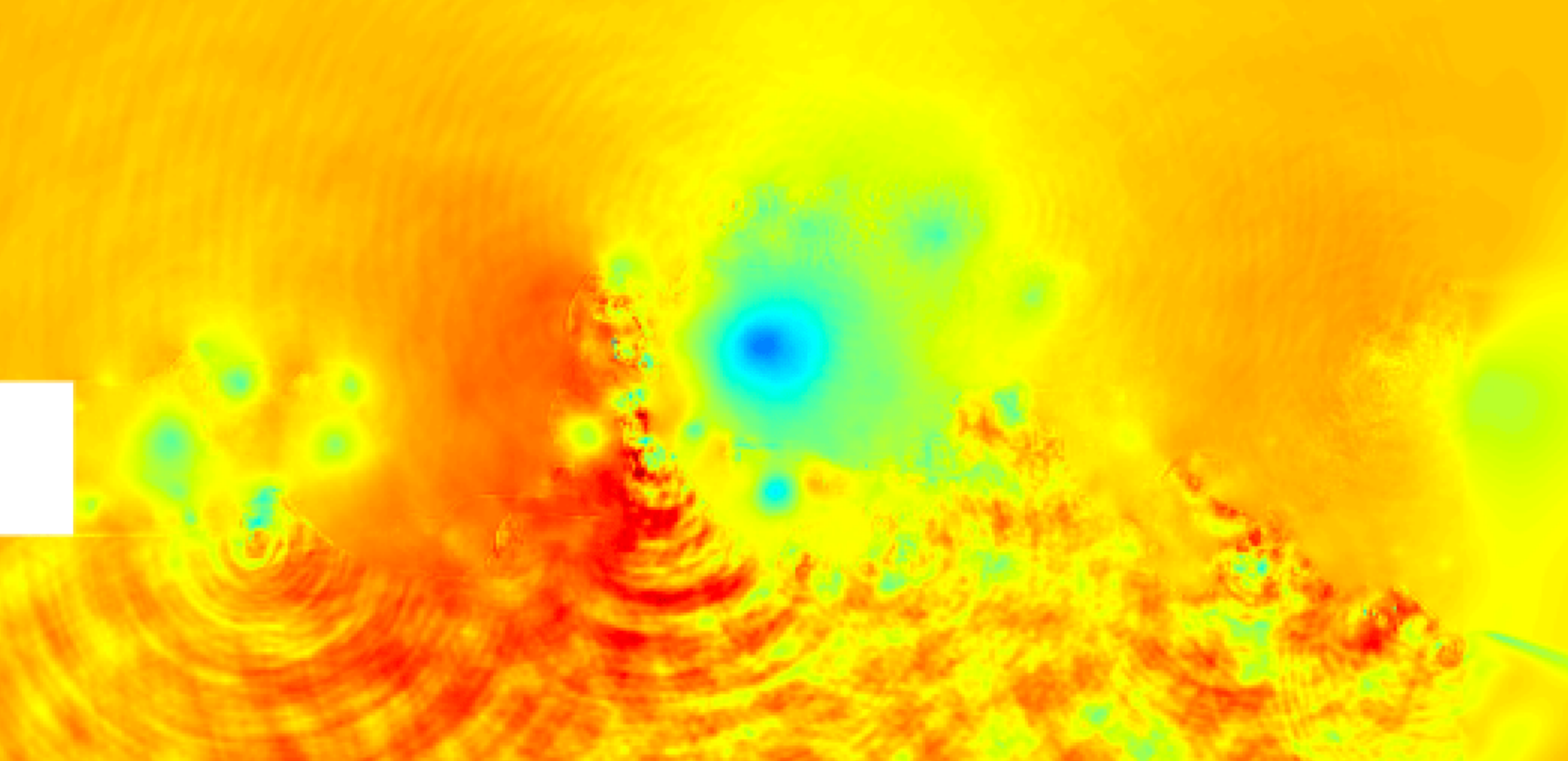}}
        {1.5pt}{1.5pt}
        \label{fig:subfig1} 
    }
    \subfigure{
        \topinset
        {\includegraphics[height=0.04\columnwidth,clip=]{P_legend.pdf}}
        {\includegraphics[width=0.475\columnwidth,trim={1cm 6cm 8.5cm 4cm},clip=]{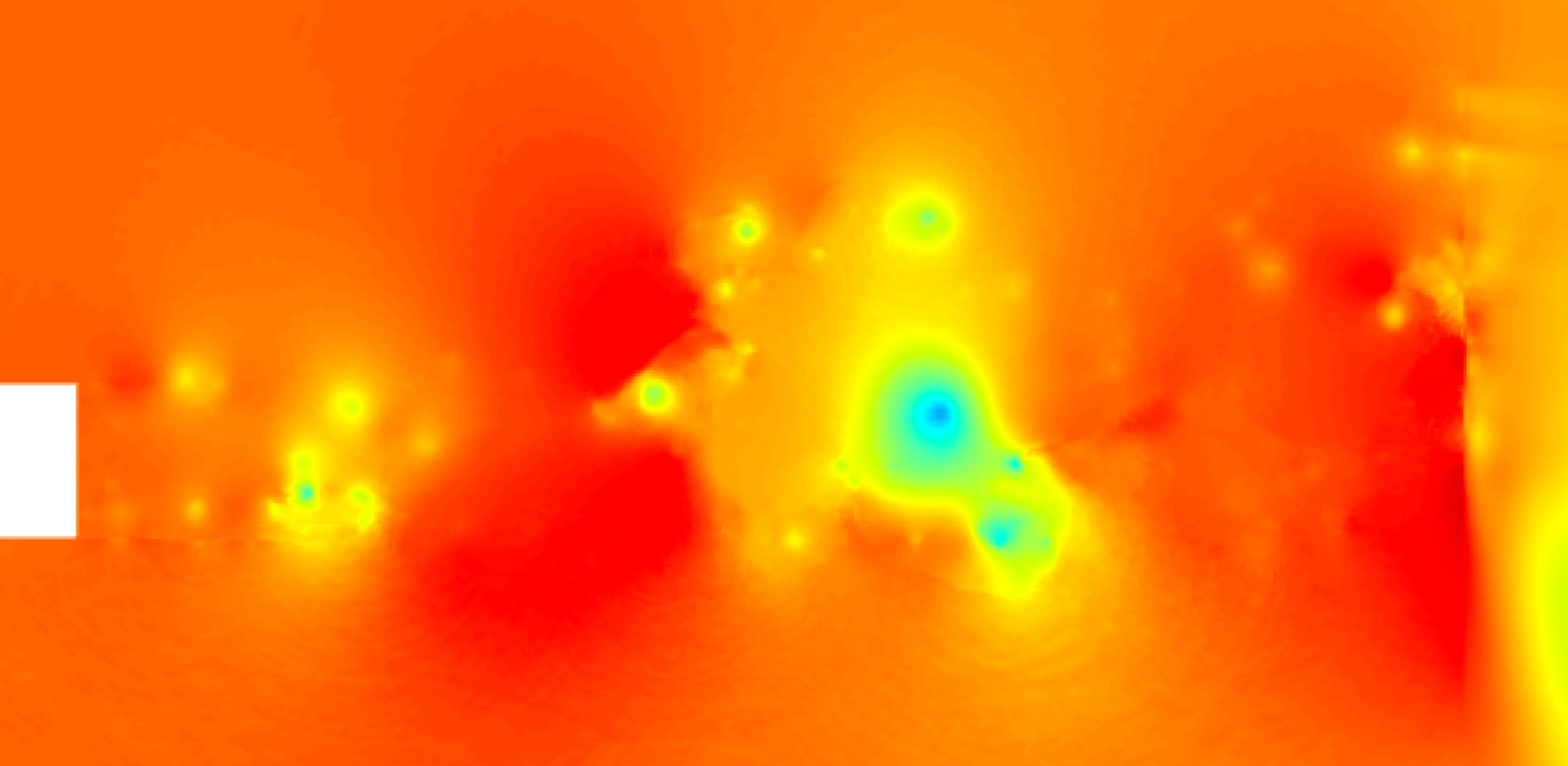}}
        {1.5pt}{1.5pt}
        \label{fig:subfig1} 
        \label{fig:subfig1} 
    }\\[-0.1em]
    % mass fraction    
    \subfigure{
        \topinset
        {\includegraphics[height=0.04\columnwidth,clip=]{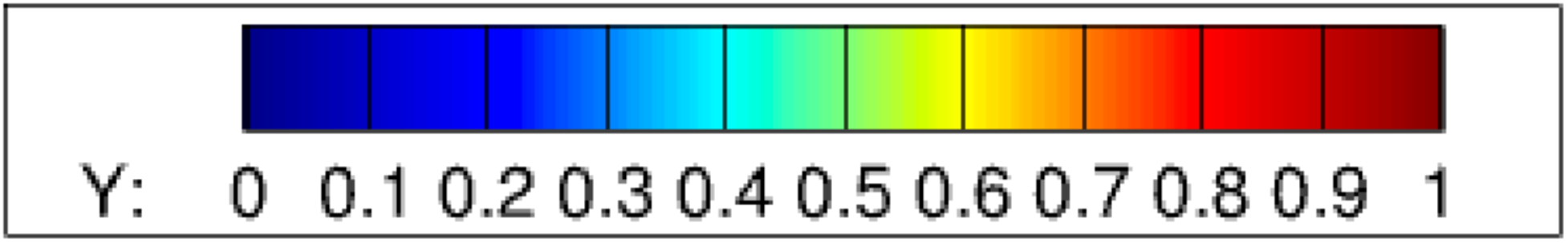}}
        {\includegraphics[width=0.475\columnwidth,trim={1cm 6cm 8.5cm 4cm},clip=]{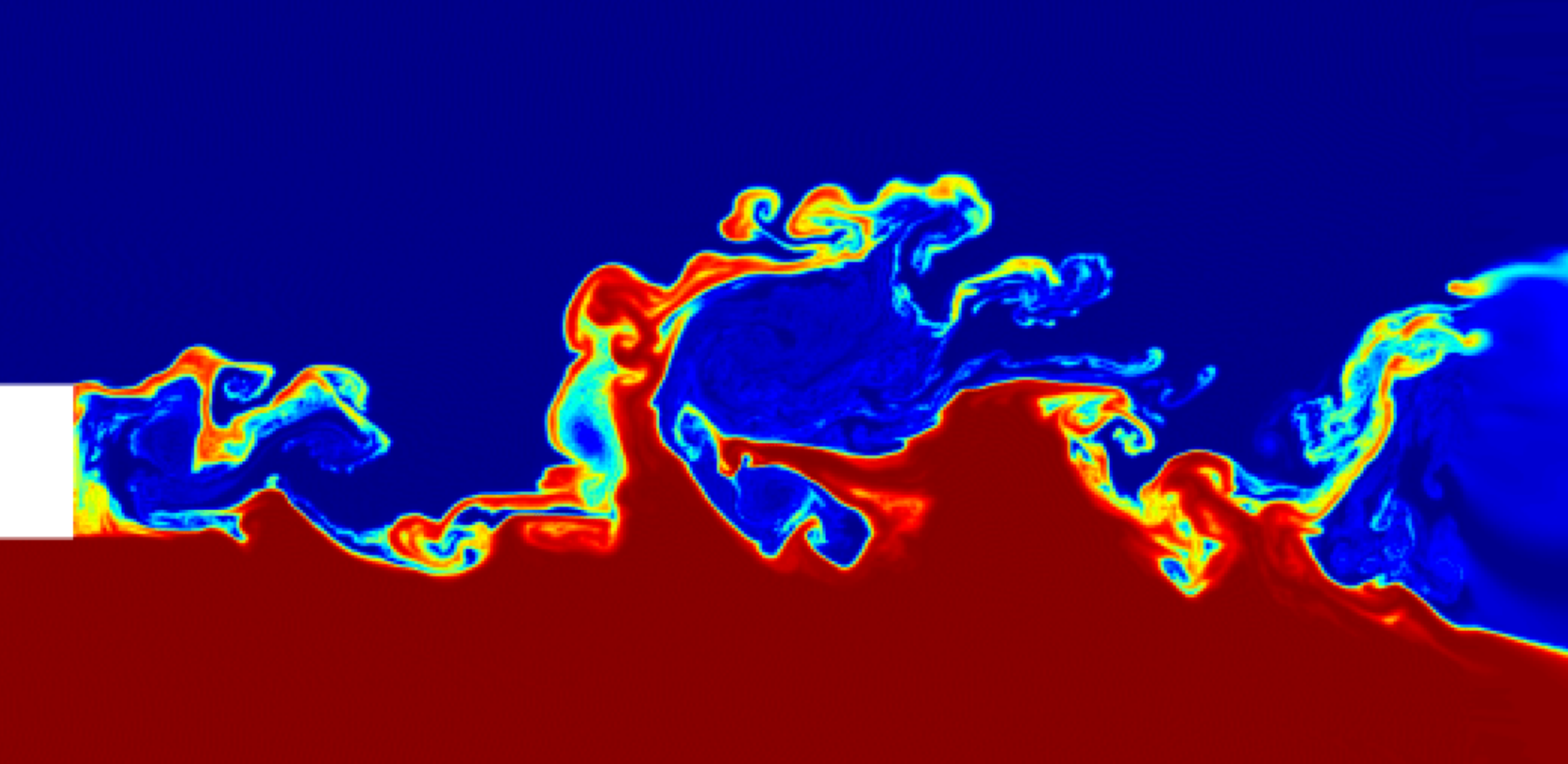}}
        {1.5pt}{1.5pt}
        \label{fig:subfig1} 
    }
    \subfigure{
        \topinset
        {\includegraphics[height=0.04\columnwidth,clip=]{Y_legend.pdf}}
        {\includegraphics[width=0.475\columnwidth,trim={1cm 6cm 8.5cm 4cm},clip=]{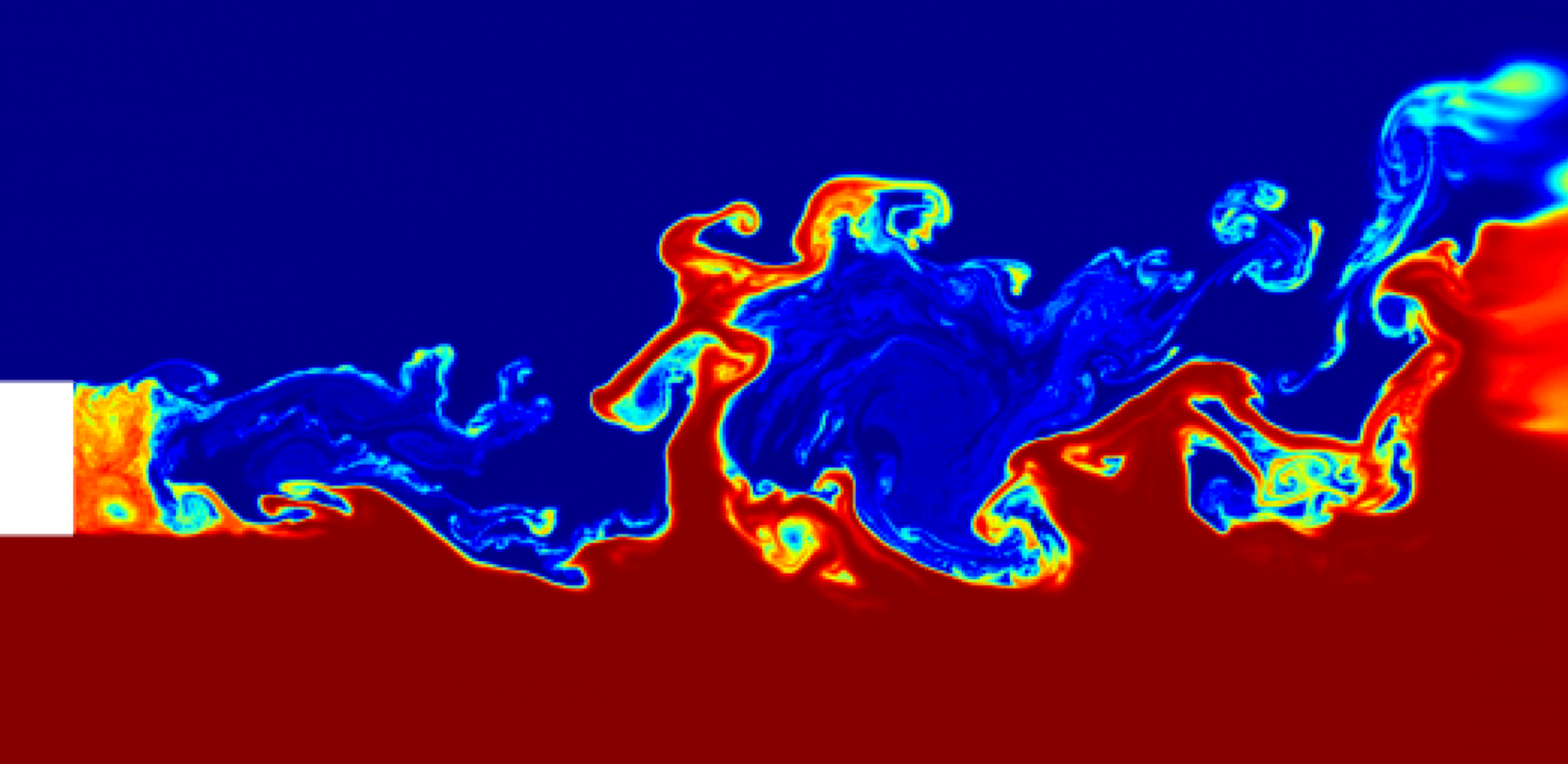}}
        {1.5pt}{1.5pt}
        \label{fig:subfig1} 
        \label{fig:subfig1} 
    }\\[-0.1em]
    % temperature    
    \setcounter{subfigure}{0}
    \subfigure[Fully conservative]{
        \topinset
        {\includegraphics[height=0.04\columnwidth,clip=]{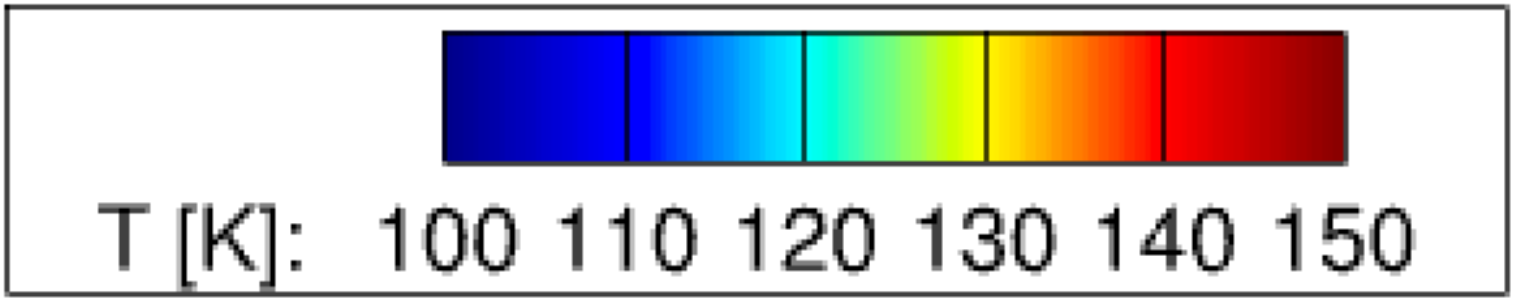}}
        {\includegraphics[width=0.475\columnwidth,trim={1cm 6cm 8.5cm 4cm},clip=]{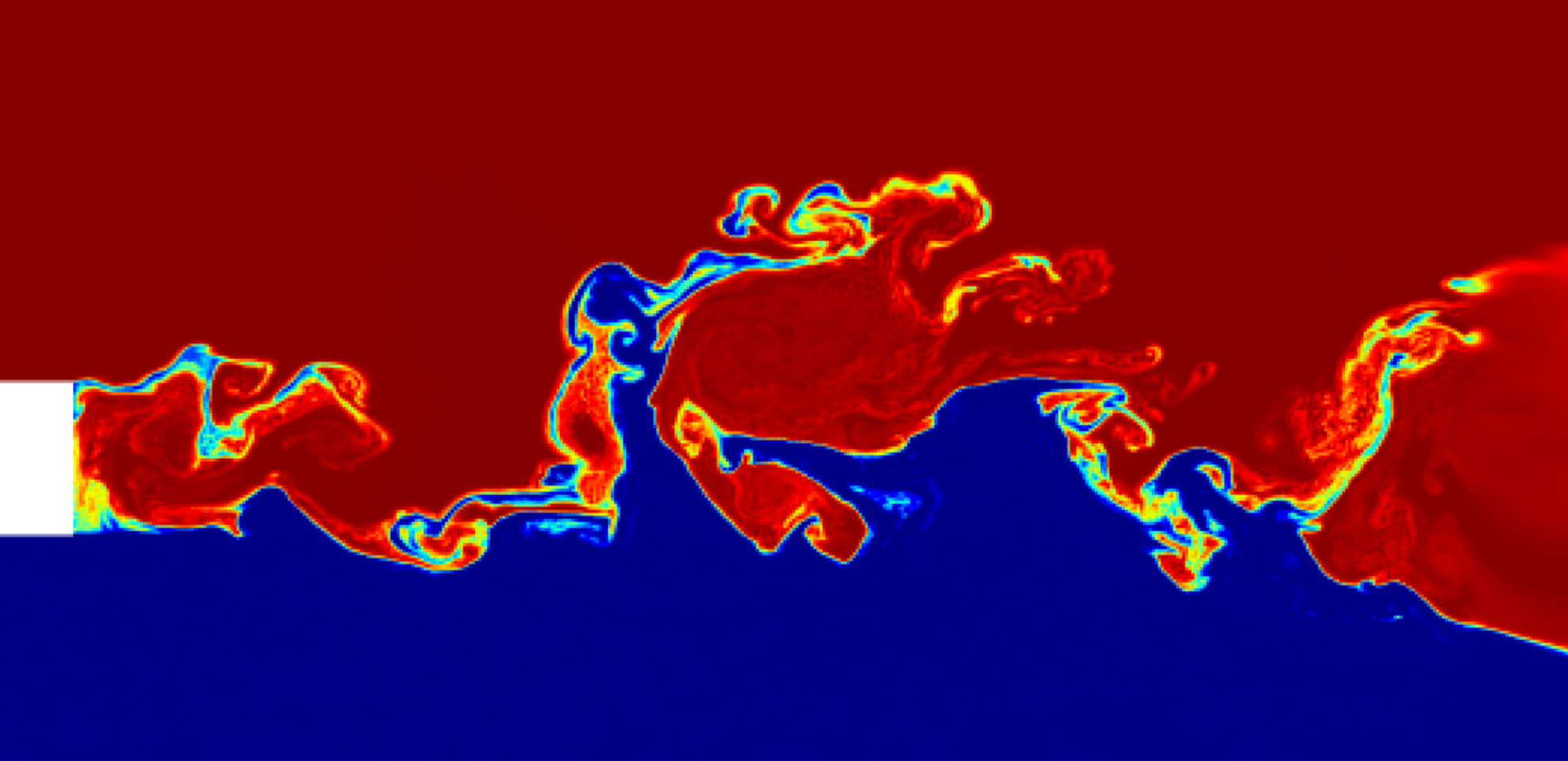}}
        {1.5pt}{1.5pt}
        \label{fig:subfig1} 
    }
    \subfigure[Double-flux model]{
        \topinset
        {\includegraphics[height=0.04\columnwidth,clip=]{T_legend.pdf}}
        {\includegraphics[width=0.475\columnwidth,trim={1cm 6cm 8.5cm 4cm},clip=]{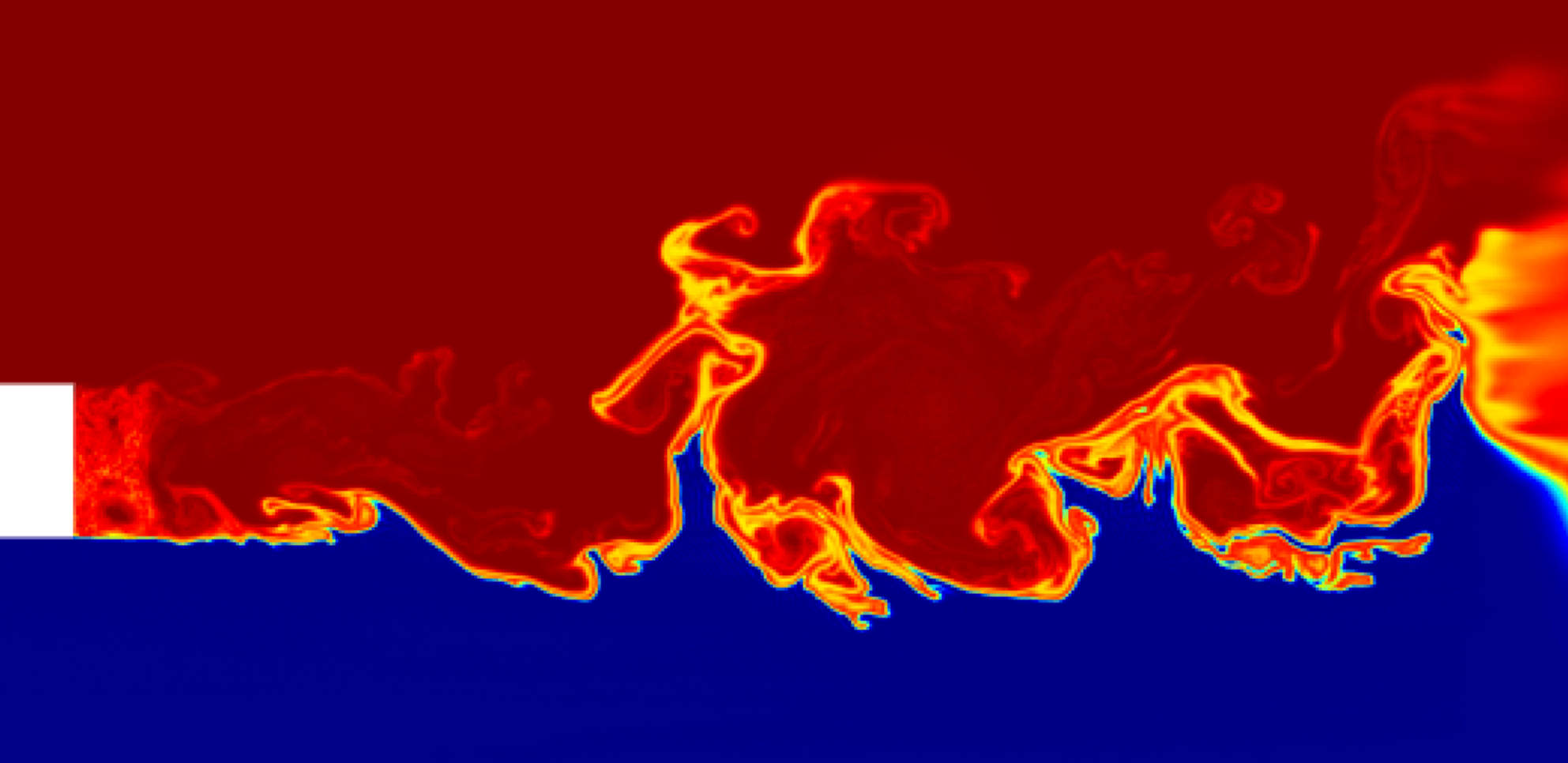}}
        {1.5pt}{1.5pt}
        \label{fig:subfig1} 
        \label{fig:subfig1} 
    }\\[-0.1em]
    \caption{Instantaneous fields of density, pressure, O$_2$ mass fraction, and temperature from top to bottom for the fully conservative scheme (left) and quasi-conservative scheme with double-flux model (right) for the LOX/GH2 mixing case.\label{fig:2DMixing_results}}
\end{figure}
%-----------------------------------------------------------------

The two streams are separated by the injector lip, which is also included in the computational domain. A domain of $15h \times 10h$ is used, where $h = 0.5$~mm is the height of the injector lip. The region of interest extends from 0 to $10h$ in the axial direction with the origin set at the center of the lip face. A sponge layer of length 5$h$ at the end of the domain is included to absorb the acoustic waves. The computational mesh has 100 grid points across the injector lip, which is fine enough to give statistically converged solution~\cite{ruiz2015numerical}. A uniform mesh is used in both directions for the region from 0 to 10$h$ in axial direction and from -1.5$h$ to 1.5$h$ in transverse direction; stretching is applied with a ratio of 1.02 only in the transverse direction outside this region. Adiabatic no-slip wall conditions are applied at the injector lip and adiabatic slip wall conditions are applied for the top and bottom boundaries of the domain. A 1/7th power law for velocity is used for both the LOX and GH2 streams. The CFL number is set to 0.8 and no subgrid scale model is used. Simulations are conducted with both the FC scheme and the QC scheme with double-flux model. Results obtained with the adaptive double-flux algorithm give similar results as in the previous subsection, and are therefore omitted here.

%-----------------------------------------------------------------
\begin{figure}[!tb!]
    \centering
    \subfigure{
        \includegraphics[width=0.48\columnwidth,clip=]{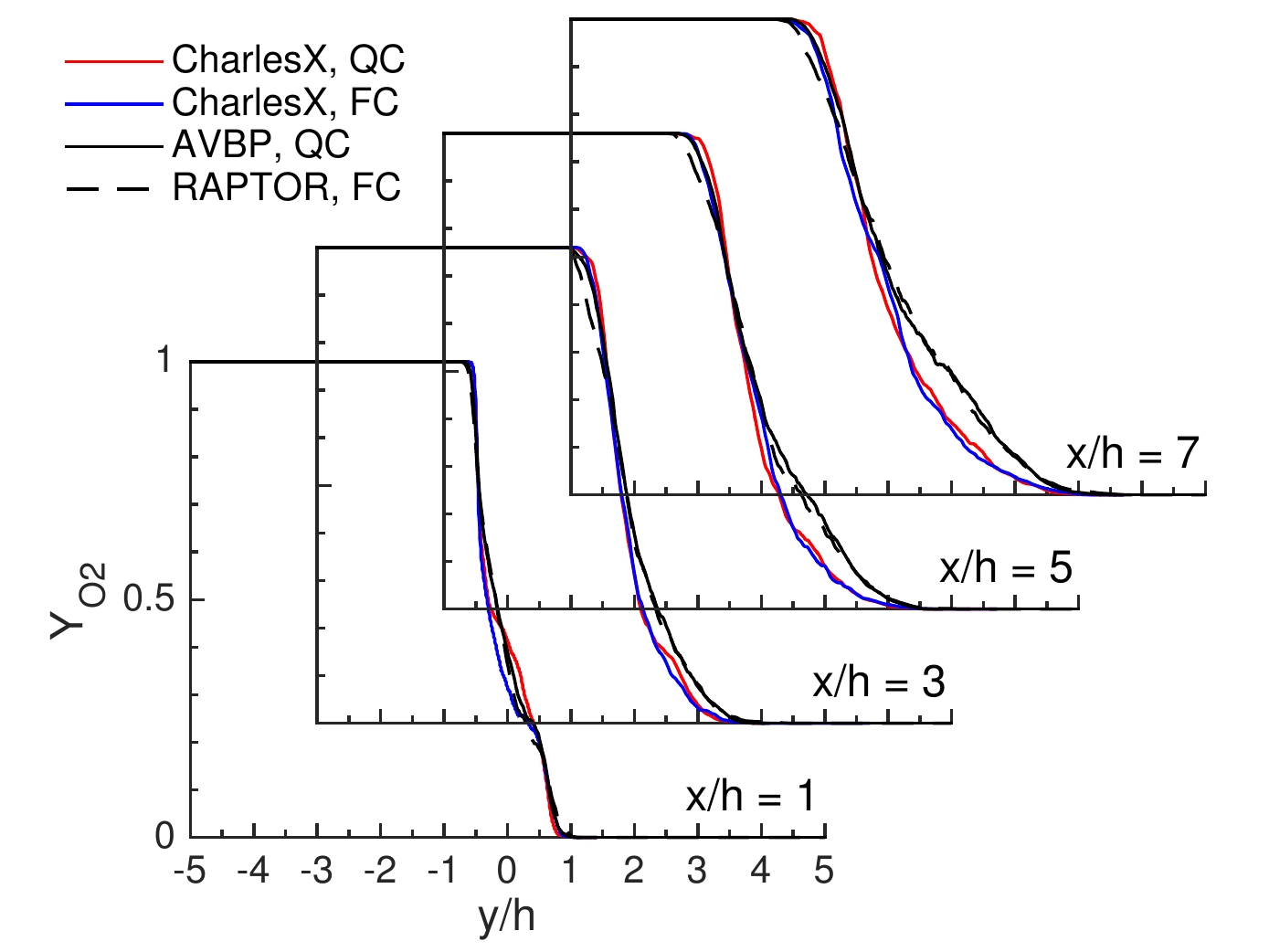}
        \label{fig:subfig1} 
    }
    \subfigure{
        \includegraphics[width=0.48\columnwidth,clip=]{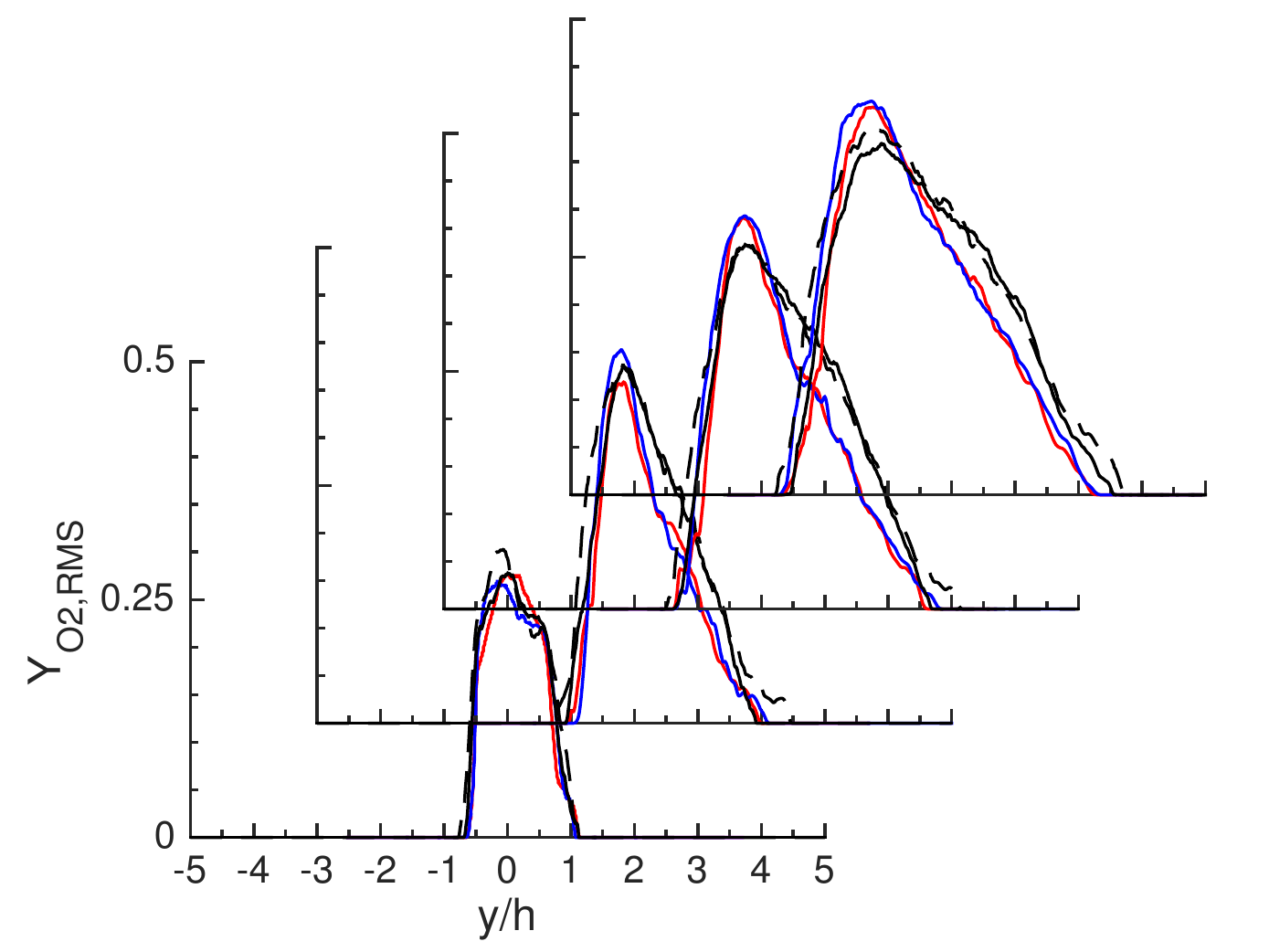}
        \label{fig:subfig1} 
    }
    \caption{Transverse cuts of (a) mean and (b) rms oxygen mass fraction of the LOX/GH2 mixing case. FC is short for fully conservative and QC for quasi-conservative.\label{fig:2DMixing_YO2}}
\end{figure}
%-----------------------------------------------------------------

%-----------------------------------------------------------------
\begin{figure}[!tb!]
    \centering
    \subfigure{
        \includegraphics[width=0.48\columnwidth,clip=]{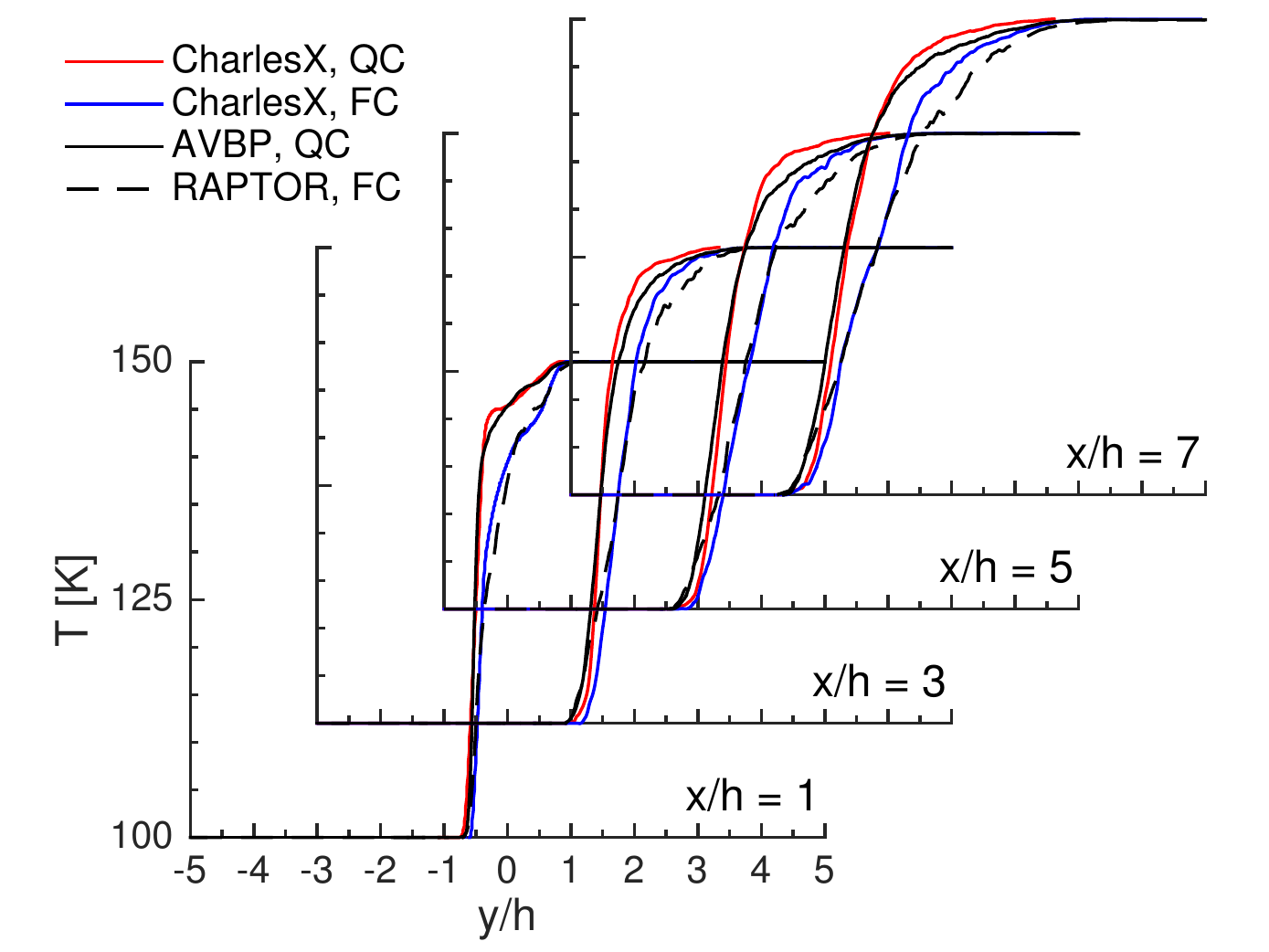}
        \label{fig:subfig1} 
    }
    \subfigure{
        \includegraphics[width=0.48\columnwidth,clip=]{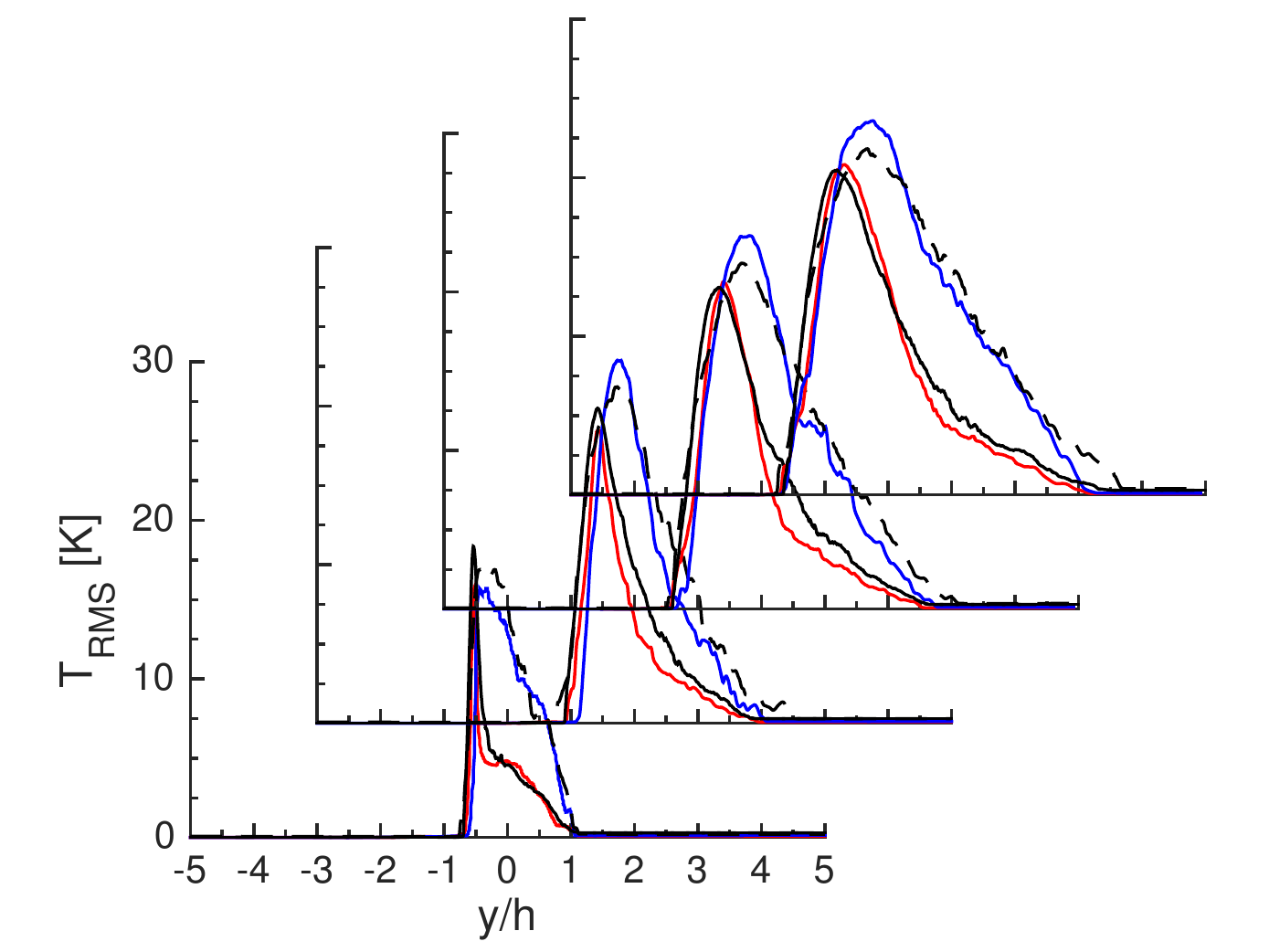}
        \label{fig:subfig1} 
    }
    \caption{Transverse cuts of (a) mean and (b) rms temperature of the LOX/GH2 mixing case. FC is short for fully conservative and QC for quasi-conservative.\label{fig:2DMixing_T}}
\end{figure}
%-----------------------------------------------------------------

% discussion on flow field and mixing behaviors
\Cref{fig:2DMixing_results} shows results for instantaneous fields of density, pressure, O$_2$ mass fraction and temperature computed by both schemes. 
%Similar results can be seen in comparison with the {\it n}-dodecane injection case in the previous subsection. 
For the density field, similar ``comb-like" structures~\cite{mayer2000injection, ruiz2012unsteady} can be seen for both schemes. This was also observed experimentally under typical rocket engine conditions~\cite{mayer2000injection, oschwald2006injection, chehroudi2012recent}. Spurious pressure oscillations can clearly be seen with the FC scheme. The oscillations are generated mostly from the oxygen side due to the fact that the pseudo-boiling process is undergone with almost pure oxygen~\cite{banuti2017seven}. Despite the similar behavior in the mass fraction field, the temperature field shows surprisingly distinct behavior. The temperature field predicted by the double-flux model is significantly higher than that predicted by the FC scheme.

% comparison with other solvers
The flow field was averaged over 15 flow-through-times after reaching steady state, with one flow through time corresponding to 0.125~ms~\cite{ruiz2015numerical}. Mean and rms results for O$_2$ mass fraction and temperature are shown in \cref{fig:2DMixing_YO2,fig:2DMixing_T}. Statistics at different axial locations ($x/h$~=~1,~3,~5,~7) are plotted as a function of normalized transverse distance. The results obtained in the benchmark case~\cite{ruiz2015numerical} are also included for comparison, which was computed using the solvers AVBP and RAPTOR. AVBP uses a QC scheme for the simulation of transcritical flows, whereas RAPTOR is a FC solver with a dual time-stepping scheme. \Cref{fig:2DMixing_YO2} show that, for O$_2$ mass fraction, there is a good agreement between the FC and QC schemes with the solver CharLES$^x$, and the mean and rms values collapse between the two schemes. Similar trends can be observed for the other two solvers. However, results obtained from CharLES$^x$ show slight discrepancies from the other two solvers, which is reflected by a slightly narrower shear layer on the hydrogen side. These discrepancies can be attributed to the different implementations of the sponge layer and outlet boundary conditions adopted by the different solvers~\cite{lacaze2016private}.

Considering the mean and rms results for the temperature field shown in \cref{fig:2DMixing_T}, it can clearly be seen that two distinct behaviors are present; one obtained by the FC schemes and the other by the QC schemes. Similar to the trends seen in \cref{fig:2DMixing_results}, the mean temperature obtained by the QC schemes is significantly higher than that obtained by FC schemes and the temperature mixing layer is much narrower, which is more clearly seen by the rms results. Note that these results are obtained from three different numerical solvers, and although in this study the numerics within CharLES$^x$ is kept the same as much as possible for FC and QC schemes, the numerics from the other two solvers differ significantly. Moreover, the QC schemes used by AVBP and CharLES$^x$ are distinctly different. The surprisingly similar behavior in the temperature field observed here is solely determined by whether the numerical solver is fully conservative or not.

%-----------------------------------------------------------------
\begin{figure}[!tb!]
    \centering
    \includegraphics[height=0.36\columnwidth,trim={25 25 25 25},clip=]{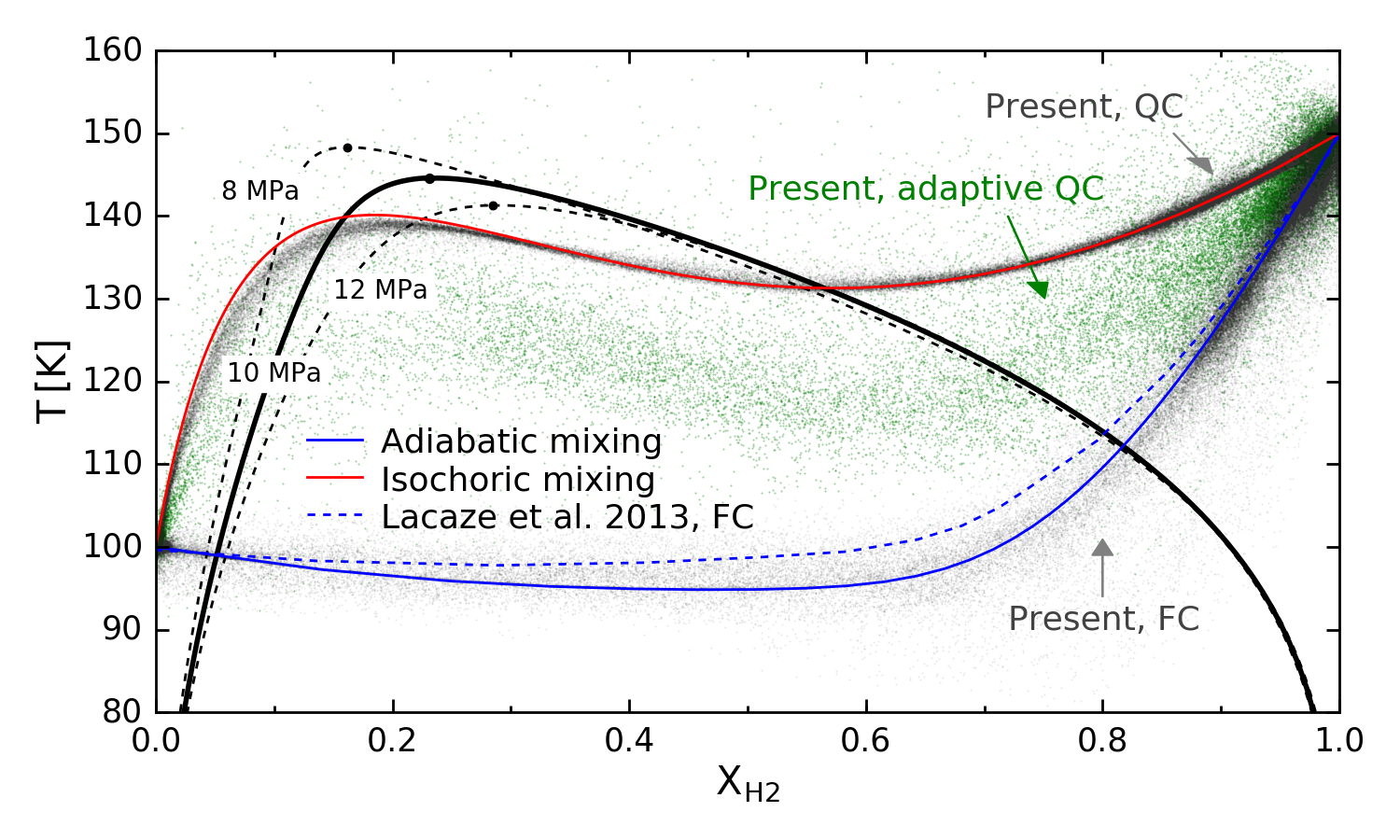}
    \caption{$T$-$X$ diagram for the LOX/GH2 mixing layer case. Scattered data from simulation results in \cref{sec:H2O2} for two different numerical schemes. FC is short for fully conservative and QC for quasi-conservative. Adaptive QC refers to the QC scheme with adaptive double-flux model ($\epsilon_{\gamma^*} = 8.0$). Phase boundaries from VLE calculations are shown in black for three pressures. Simulation results from Lacaze~et~al.~\cite{lacaze2013modeling} (FC scheme) are also shown for reference.\label{fig:Tx_H2O2}}
\end{figure}
%-----------------------------------------------------------------

% scatter plot
The instantaneous simulation results are plotted as scattered data in the form of $T$-$X$ diagrams in \cref{fig:Tx_H2O2} along with the simulation results from Lacaze~et~al.~\cite{lacaze2013modeling}. 
%Results for the LOX/GH2 mixing layer case follow similar trends as seen in \cref{fig:Tx_dode} regarding the mixing lines. 
The FC schemes from both Lacaze~et~al.~\cite{lacaze2013modeling} and the present work predict a substantially lower temperature with an approximately isothermal mixing process on the LOX side compared to the sharp increase of temperature predicted by the QC scheme. The adaptive QC scheme predicts a mixing line in between the ones obtained from the other two schemes. In terms of the phase behaviors, due to the lower temperature on the hydrogen side, all three schemes predicts states within the two-phase region, although the QC scheme has the smallest percentage of solutions with phase separation.

In summary, distinct mixing behaviors are observed for FC and QC schemes. Specifically, the temperature fields predicted by QC schemes are significantly higher than that obtained from FC schemes, which results in significant difference in phase separation predictions. These differences in the mixing behaviors have significant influence on the subsequent processes after mixing, e.g. ignition and combustion~\cite{esclapez2017large, ma2017flamelet, Wang2017, yang2017global}.

%=========================================================================
\subsection{Adiabatic and isochoric mixing\label{sec:mixing}}
\begin{figure}[!b!]
    \centering
    \subfigure[Isobaric-adiabatic mixing]{
        \includegraphics[width=0.44\columnwidth,trim={180 210 90 210},clip=]{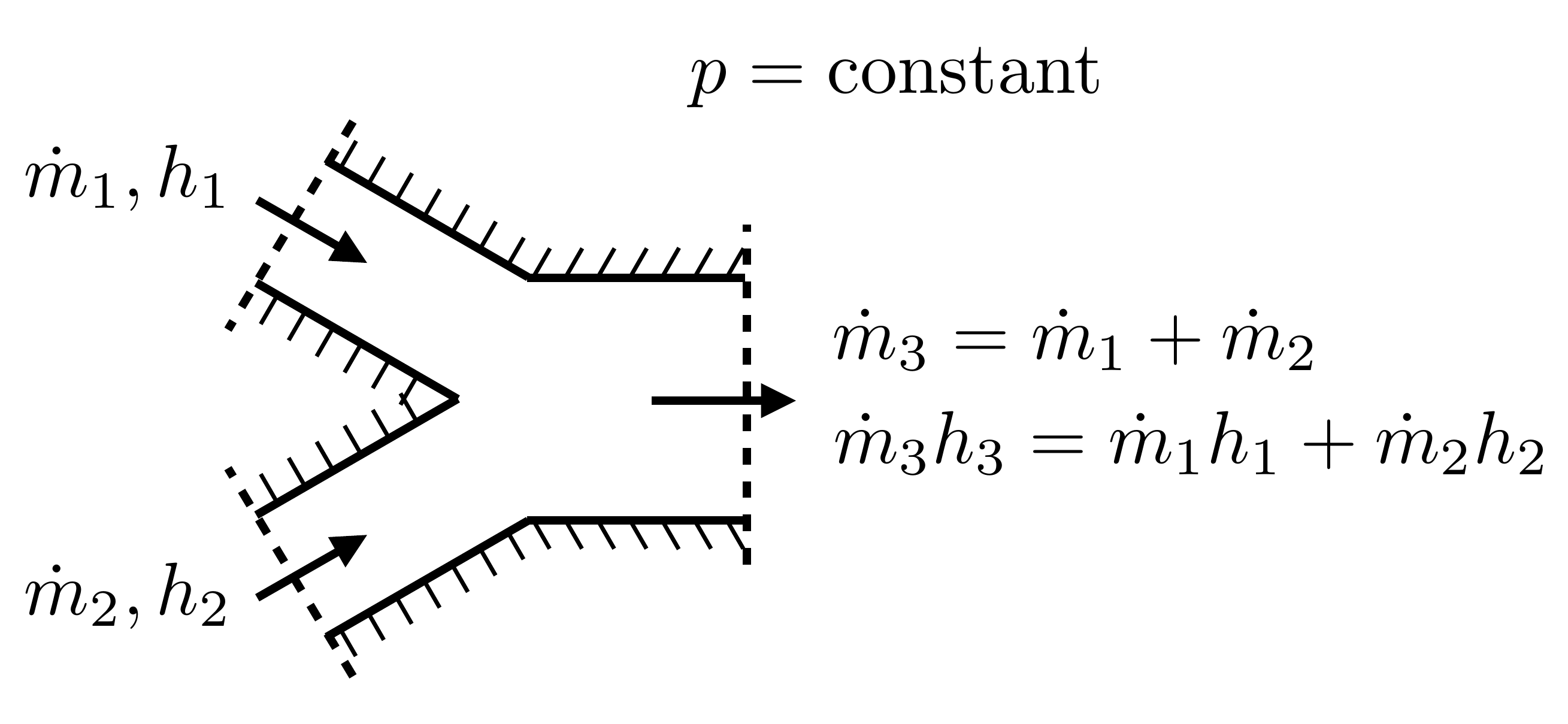}
        \label{fig:schematic_adiabatic} 
    }
    \quad
    \subfigure[Isobaric-isochoric mixing]{
        \includegraphics[width=0.44\columnwidth,trim={180 210 90 210},clip=]{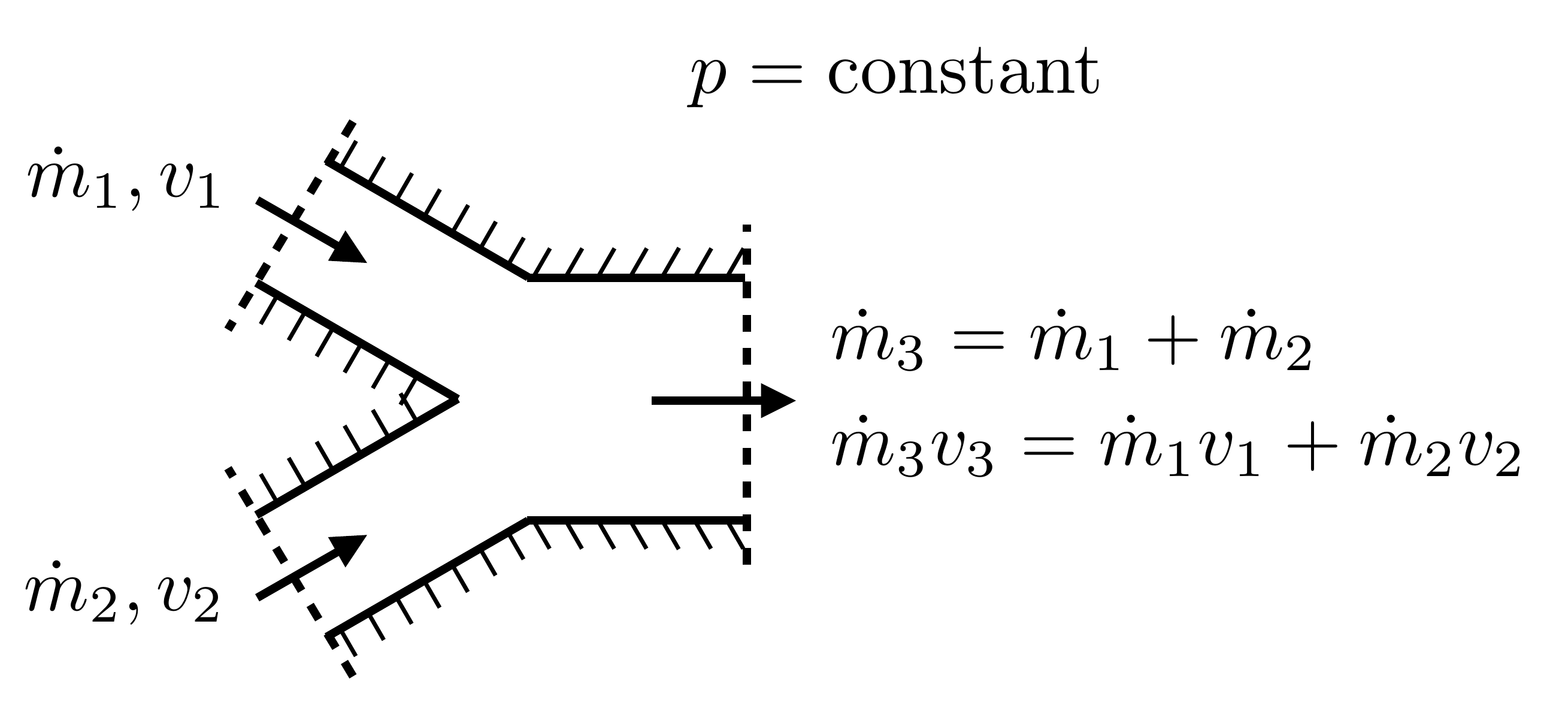}
        \label{fig:schematic_isochoric}
    }
    \caption{Schematic of (a) isobaric-adiabatic and (b) isobaric-isochoric mixing. $\dot{m}$ is the mass flow rate, $h$ is the specific enthalpy, $v$ is the specific volume, and $p$ is the pressure.\label{fig:schematic_mixing}}
\end{figure}

%=========================================================================
% two mixing models
The distinct mixing behaviors between the FC and QC schemes are found to follow two asymptotic mixing models for binary mixtures, which are the isobaric-adiabatic mixing and the isobaric-isochoric mixing models. \Cref{fig:schematic_mixing} gives examples for these two mixing models. For most applications relevant for this study, the thermodynamic pressure of the system is maintained to be constant, e.g. by the choked nozzle in rocket engines, or during the isobaric process at the top-dead-center in diesel engines, and therefore, isobaric mixing processes are considered. If the two streams are mixed adiabatically as shown in \cref{fig:schematic_adiabatic}, no heat exchange is present, and the enthalpy is conserved. Therefore, we have
\begin{subequations}
    \begin{align}
        &\dot{m}_3 = \dot{m}_1 + \dot{m}_2  \;,\\
        &\dot{m}_3 h_3 = \dot{m}_1 h_1 + \dot{m}_2 h_2\;,
    \end{align}
\end{subequations}
where $\dot{m}$ is the mass flow rate of the stream and $h$ is the specific enthalpy. Then the specific enthalpy is a linear function of the mass fraction of either stream,
\begin{equation}
    \label{eqn:adiab}
    h_3 = Y_1 h_1 + (1-Y_1) h_2 = Y_1 h_1 + Y_2 h_2\,,
\end{equation}
with $Y_1 = \dot{m}_1 / (\dot{m}_1 + \dot{m}_2) = \dot{m}_1 / \dot{m}_3$ and $Y_2 = \dot{m}_2 / (\dot{m}_1 + \dot{m}_2) = \dot{m}_2 / \dot{m}_3$. Such isenthalpic irreversible mixing processes occur in many industrial applications. The adiabatic mixing model has been used in several studies related to transcritical flows~\cite{Dahms2013a, lacaze2015analysis, matheis2016multi, Qiu2015} either for the validation and comparison with numerical simulation results, or as an assumption for the analysis of phase separation behaviors.

%-----------------------------------------------------------------
%-----------------------------------------------------------------

In contrast to the adiabatic mixing, if the two streams are mixed isochorically as shown in \cref{fig:schematic_isochoric}, the volume of the binary mixture is conserved,
\begin{subequations}
    \begin{align}
        &\dot{m}_3 = \dot{m}_1 + \dot{m}_2  \;,\\
        &\dot{m}_3 v_3 = \dot{m}_1 v_1 + \dot{m}_2 v_2\;,
    \end{align}
\end{subequations}
and this gives a specific volume profile linear in the mass fraction,
\begin{equation}
    \label{eqn:isoch}
    v_3 = Y_1 v_1 + (1-Y_1) v_2 = Y_1 v_1 + Y_2 v_2\,.
\end{equation}
This process is achieved by heat transfer between the mixture and the surroundings.

For a given constant pressure, the adiabatic mixing line in the $T$-$X$ state space can be computed from \cref{eqn:adiab} by converting enthalpy values into corresponding temperature values. Similarly, the isochoric mixing line can be computed from \cref{eqn:isoch} by calculating temperature from specific volume. The mixing lines from these two mixing models are calculated with the PR-EoS under conditions of test cases in \cref{sec:H2O2}, in which the constant environmental pressure is specified using the nominal chamber pressure. The results are plotted in \cref{fig:Tx_H2O2} along with corresponding simulation solutions. It is interesting to see that for the LOX/GH2 mixing layer cases, simulation solutions from FC schemes closely follow the adiabatic mixing model, and simulation solutions from QC schemes almost collapse with the isochoric mixing model. This is also true for the simulation results from other studies~\cite{lacaze2013modeling} shown in \cref{fig:Tx_H2O2}.

%=====================================================================
\section{Conclusions}\label{sec:conclusions}
%=====================================================================
The impact of different numerical schemes of commonly used diffused interface methods on the mixing processes for transcritical flows is investigated in this study. To this end, two classes of schemes are examined, namely the fully conservative (FC) and quasi-conservative (QC) schemes. An adaptive double-flux model, which is a hybrid of the FC and QC schemes, is developed using a sensor that is based on the gradient of the adiabatic exponent.

Significant spurious pressure oscillations are observed in multi-dimensional test cases for the FC scheme, which is consistent with the conclusions of previous studies. Distinctly different mixing behaviors are observed for FC and QC schemes, especially for the temperature field. It was found that for underresolved solutions, the FC and QC schemes behave as the isobaric-adiabatic and isobaric-isochoric mixing models for binary mixtures. The adaptive QC scheme has a behavior in between the two. These are confirmed by numerical simulation results.

The analysis of this study provides insights into the interpretation of LES results of transcritical flows. More experimental and computational investigations are needed to investigate the real physical mixing and phase separation behaviors of transcritical flows.

%=====================================================================
\section*{Acknowledgments}
%=====================================================================
Financial support through NASA with award numbers NNX14CM43P and NNM13AA11G are gratefully acknowledged. Part of this work was inspired by the discussion with Dr. Jan Matheis and Prof. Stefan Hickel during the Summer Program 2016. The authors would also like to thank Dr. Luis Bravo for his help on this paper. Resources supporting this work were provided by the NASA High-End Computing (HEC) Program through the NASA Advanced Supercomputing (NAS) Division at Ames Research Center.

%=====================================================================
%\section*{References}
%=====================================================================
\bibliographystyle{aiaa}
\bibliography{references}

\end{document}